# Backbone $N_xH$ Compounds at High Pressures

Alexander F. Goncharov [1,2], Nicholas Holtgrewe[2,3], Guangrui Qian [4,5,7], Chaohao Hu[7], Artem R. Oganov [4,5,6,7], M. Somayazulu[2], E. Stavrou[2], Chris J. Pickard[8], Adam Berlie[1], Fei Yen[1], M. Mahmood[3], S. S. Lobanov[2,9], Z. Konôpková[10], V. B. Prakapenka[11]

[1]Key Laboratory of Materials Physics, Institute of Solid State Physics, Chinese Academy of Sciences, 350 Shushanghu Road, Hefei, Anhui 230031, China

[2]Geophysical Laboratory, Carnegie Institution of Washington, 5251 Broad Branch Road, Washington, D.C. 20015, USA

[3]Howard University, Washington, DC 20059, USA

[4] Department of Geosciences, State University of New York, Stony Brook, NY 11794-2100

[5] Center for Materials by Design, Institute for Advanced Computational Science, State University of New York, Stony Brook, NY 11794-2100

[6]Moscow Institute of Physics and Technology, 9 Institutskiy lane, Dolgoprudny city, Moscow Region, 141700, Russian Federation

[7] Guangxi Key Laboratory of Information Materials, Guilin University of Electronic Technology, Guilin 541004, China

[8]University College London, Gower St, London, WC1E 6BT, U.K.

[9]V.S. Sobolev Institute of Geology and Mineralogy, SB RAS, 3 Pr. Ac. Koptyga, Novosibirsk 630090, Russia.

[10] DESY Photon Science, Notkestrasse 85, D-22607, Hamburg, Germany.

[11]Center for Advanced Radiation Sources, University of Chicago, Chicago, IL  60637, USA

**Optical and synchrotron x-ray diffraction diamond anvil cell experiments have been combined with first principles theoretical structure predictions to investigate mixed $N_2$ and $H_2$ up to 55 GPa. We found the formation of oligomeric $N_xH$ ($x \geq 1$) compounds using mechano- and photochemistry at pressures above 47 and 10 GPa, respectively, and room temperature. These compounds can be recovered to ambient pressure at T<130 K, whereas at room temperature, they can be metastable down to 3.5 GPa. Our results suggest new pathways for synthesis of environmentally benign high energy-density materials and alternative planetary ice.**

Chemistry of the N-H system is quite rich resulting in a number of known compounds that include ammonia, hydrazine, tetrazene, and hydrazinium or ammonium azides. However, only ammonia is the stable compound at ambient pressure, while others that have N-N bonds are metastable. Accordingly, the planetary models assumed that ammonia would constitute a substantial amount of interior of giant planets such as Uranus and Neptune [1]. On the other hand, metastable N-H compounds (as well other nitrogen rich systems) at ambient pressure would be superior high energy-density materials (HEDM) due to the substantial energy difference between the single and triple nitrogen-nitrogen bonds (*e.g.*, [2]) (167 vs 942 kJ/mole, respectively). Additionally, such materials have much lower environmental and safety hazards compared to conventional energetic materials as their primary decomposition product is molecular nitrogen ($N_2$) and they have higher thermal stability. Although all-nitrogen material (*e.g.*, cubic gauche nitrogen, cgN [3]) would be an ultimate HEDM, its metastability at close to ambient conditions remains problematic. Moreover, synthesis of such materials requires very high pressures and temperatures, making it unpractical. Large polynitrogen molecules have been theoretically predicted but experimental realization has not been demonstrated [4, 5].

While synthesis of pure polymeric nitrogen at low pressures remains elusive, polymeric nitrogen-rich mixed materials would be a great alternative to commonly used explosives. There are numerous reports on synthesis of complex energetic materials containing chains of nitrogen atoms: N5, N8, and even N10 [6-8] and complex "salts" such as TAG-MNT [9], but these materials require other elements for stabilization and their synthesis remains quite challenging. Rich in nitrogen N-H systems appear to be the most attractive as they would be the most energy-density efficient and because hydrogen tends to stabilize low bond order nitrogen compounds (*e.g*., hydrazine).

Application of high pressure provides an alternative to pure chemical routes for making new N-H materials as it stabilizes distinctive bonding schemes, such as single N-N bonds [10]. Indeed, recent theoretical work using *ab initio* evolutionary structure search predicted a new polymeric $(NH)_4$ hydronitrogen solid, which becomes more stable than ammonium azide and *trans*-tetrazene at 36 and 75 GPa, respectively [11]. However, using only thermodynamic stimuli is often insufficient to create new materials because of large kinetic barriers between different bonding schemes, which may require high temperature [3] or radiation [12-14] to create the necessary reaction conditions.

Previous experimental works on high-pressure behavior of $N_2$-$H_2$ mixtures found anomalous behavior of the $N_2$ and $H_2$ Raman vibron modes compared to pure materials [15, 16]. Ciezak *et al*. [15] reported coexistence of two unidentified solid phases above 35 GPa, one of which was suggested to be amorphous based on a broadening of the Raman spectra and change in color. However, no definite conclusion about the chemical transformations could be made.

Here we show that the $N_2$-$H_2$ system at moderate pressures near 47 GPa is prone to molecular dissociation and formation of an oligomeric and/or polymeric N-H compound even at room temperature, in striking difference to the behavior of pure $N_2$ [3]. This unusual behavior is a unique example of mechanochemistry – the coupling of chemical processes and intermolecular interactions [17, 18], which have never been documented previously for simple diatomic molecules. Moreover, we find that application of a range of ultra-short pulses, from near ultraviolet (UV) (370 nm) to near infrared (IR) radiation (720 nm), causes a similar photochemical reaction even at 10 GPa. Our experimental observations are supported by the results of evolutionary and random structural search, which shows that oligomeric and/or polymeric N-H compounds of various stoichiometries become increasingly stable at high pressures. Our findings suggest that polymeric/oligomeric hydronitrogen compounds can be kinetically stable (i.e. metastable) at ambient pressure, while these materials become increasingly thermodynamically stable at high pressures suggesting that there can be planetary ices that have an alternative to common ammonia composition.

Raman spectroscopy, synchrotron X-ray diffraction (XRD), and visual observations show that the mixture of molecular $N_2$ and $H_2$ [19] remains a single-phase fluid up to 9-11 GPa and solidifies at higher pressure (see the Supplemental Material [19], Fig. S1). The solid alloy forms crystallites (~10 μm) (Fig. 1(a)); the Raman microprobe shows qualitatively similar spectra at various positions, which for some grains differ slightly in the Raman peak intensity, suggesting that the composition of crystallites can vary spatially. Given the complexity of the Raman spectra, which cannot be described as a superposition of molecular $N_2$ and $H_2$, and the presence of the low-angle Bragg peaks in X-ray diffraction (see Figs. S2 and S3 in Ref. [19]), this material can be characterized as a van-der-Waals crystal with a large unit cell (see also Ref. [16]); the crystal structure of this material will be reported elsewhere.

Above 47 GPa substantial changes in the material vibrational and structural properties have been revealed by Raman, infrared (IR), and optical spectroscopy, as well as XRD (Figs. 1, 2, and Figs.

S4 and S5 in Ref. [19]). The sample changes its appearance: it becomes yellowish as the optical bandgap develops in the visible near 2.5 eV (Fig. S4 in Ref. [19]) and the micrograin structure changes (Fig. 1(a)). Our Raman, optical, and X-ray microprobes show almost uniform properties of this material on a length scale of several micrometers (c.f. Ref. [15]). Above 47 GPa, Raman and IR observations showed time dependent responses with characteristic times of hours and even days (Fig. 1(c), Fig S5 in Ref. [19]). The most striking feature is a strong decrease in intensity of the vibron spectra of $H_2$ and $N_2$. The hydrogen vibron modes and the low-frequency rotational bands totally disappear suggesting a completion of the chemical reaction. A new system of broad bands appears at high pressures. One of these bands is slightly lower in frequency and much broader compared to the $\nu_2$ $N_2$ vibron mode (Fig. S5 in Ref. [19]); the others have very different frequencies. Strong Raman and IR modes are observed near 3350, 1680, 1080, and 1300 cm$^{-1}$ (Raman only) (Figs. 1(c,d)). Based on their frequencies, these bands can be tentatively assigned to the stretching and deformation (scissoring and rocking) N-H modes, and stretching N-N modes (*e.g.*, Ref. [20]), respectively. In addition, a broad low-frequency strongly pressure dependent mode has been observed (Fig. 1(c)), which, based on this behavior, should be assigned to lattice translational/ libration motions. X-ray diffraction patterns drastically change through the transition at 47 GPa, where narrow Bragg reflections of the mixed $N_2$-$H_2$ crystal phase disappear and are superseded by two systems of diffraction peaks: narrow and broad (Fig. 2 and Fig. S6 in Ref. [19]).

Application of pulsed radiation (370-720 nm) causes a similar transformation to occur at much lower pressure just after solidification (>10 GPa). Based on null observations from focusing continuous wave lasers (488 & 532 nm) on the sample, this is thought to be a multi-photon process, where at least two photons promote $N_2$ and/or $H_2$ molecules into electronic excited states, inducing chemical reactivity. The complete transformation was difficult to reach even after a prolonged irradiation, thus the new material coexists with unreacted sample in the high-pressure cavity (Fig. 1(b)). In contrast to the materials synthesized at high pressures (>47 GPa), that show a wide bandgap, the irradiated material appears to be opaque and there are differences in relative intensities of the Raman bands (Fig. 1(c)). Upon decompression at room temperature, the new phase stays intact down to 3.5 GPa, even though the surrounding unreacted material transitions back to liquid.

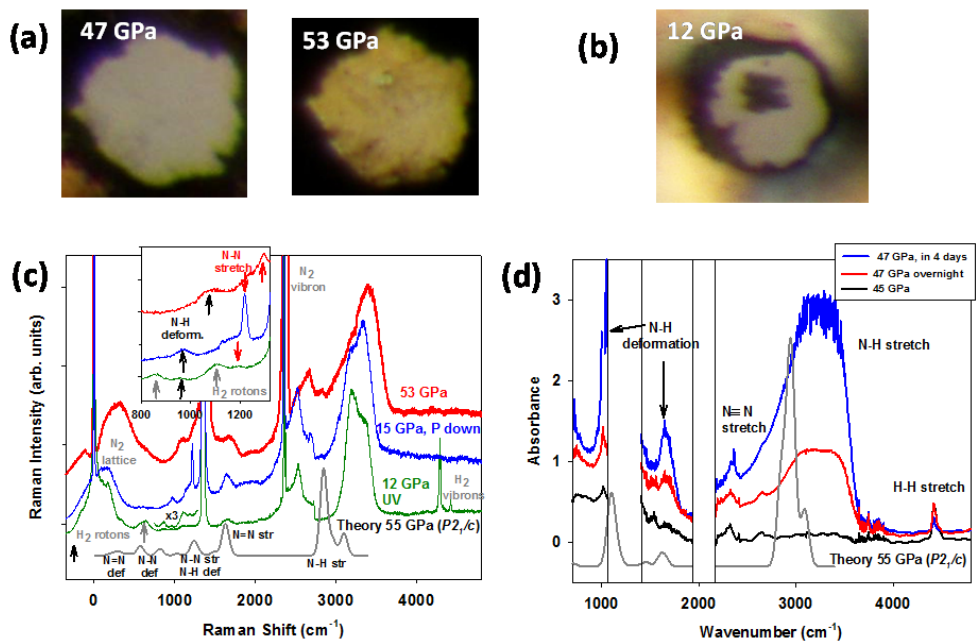

**Fig. 1 (color online). Transformation of the $H_2$-$N_2$ van-der-Waals crystal at P>47 GPa and 11.3 GPa: (a)** microphotographs showing a change in color and grain structure; **(b)** image shows the formation of a new phase (a dark spot) after UV irradiation at 11.3 GPa **(c, d)** Raman and IR absorption spectra of new synthesized materials. In (c) the two top curves correspond to the material synthesized above 47 GPa (red) and unloaded down to 15 GPa

**(blue); the bottom curve –to UV irradiated material. In (d) the IR spectra evolution with time is shown to illustrate the reaction kinetics at 47 GPa. The inset in (c) shows the details of the N-N stretch and N-H deformation mode behavior. In (d) the spectral areas of large diamond anvil absorption are masked by boxes. In (c) and (d) the experimental data are compared to the theoretically computed Raman and IR spectra of the _P2₁/c_ tetrazene – the most stable theoretically predicted structure for the NH compounds.**

Vibrational spectroscopy data provide rather tight constraints on the chemical structure of the high-pressure phase we have synthesized. Given the vibrational frequencies observed, the data clearly indicate the formation of either oligomeric or polymeric single-bonded N-N chains with the attached hydrogen atoms. This explains all the observed Raman and IR modes (Fig. 1), the absence of the $H_2$ roton and vibron modes, and the peculiar behavior of the $N_2$ vibron mode described above (Fig. S5 in Ref. [19]). This band has been tentatively assigned to guest $N_2$ molecules imbedded to the high-pressure phase matrix and/or nanosized crystalline nitrogen. Additional information about the chemical structure of the synthesized products can be obtained based on the N-N stretch vibrational frequency (Fig. 3a). We used the high pressure behavior of hydrazine ($H_2N$-$NH_2$), which has similar vibrational spectra, as a reference. We find that the N-N stretch mode frequency of the material synthesized at high pressure is somewhat higher than that of hydrazine, while the same mode of the UV irradiated product is very close in frequency to hydrazine (Fig. 3a). Based on analogy with the better studied C-H system, where the C-C stretch frequency depends on the length of the -C-C- chain [21], we suggest that the material synthesized at high pressure is polymeric or oligomeric (at least tetramers), while that synthesized using UV irradiation consists of shorter N-N chains. Moreover, we find that the material unloaded to ambient pressure at low temperatures (80 K) shows the reduced value of the N-N stretch mode frequency (Fig. 3a), which is consistent with shortening of N-N chains at lower pressures.

The Raman and IR bands of newly synthesized N-H materials are broadened compared to common molecular solids (*e.g.*, hydrazine), which suggests either large pressure gradients or it may be due to compositional or structural defects. The latter is consistent with the fact that the transformation has been accomplished at room temperature, and thus has been a subject of kinetic and or steric hindrance (*e.g.*, Ref. [22]). We propose that this could result in the stacking defects as the formation of N-N long chains can be directionally frustrated. However, the presence of narrow diffraction lines in some experiments (Fig. 2) suggests that a long range order has been formed. Our attempts to anneal the stresses and inhomogeneities using gentle laser heating (<1000 K) caused the high-pressure phase to decompose; the obvious reaction product was molecular $N_2$.

The analysis of the diffraction patterns at 55 GPa (Fig. 2 and Fig. 4S) shows that broadened XRD peaks correspond to ε-nitrogen [23] at somewhat lower pressure (45 GPa), the broadening being due to nanosized crystallites formed (Fig. S4 in Ref. [19]). This is in-line with the observation of broadened Raman $N_2$ vibron and a low-frequency lattice mode (Fig. 1). The remaining (narrow) diffraction peaks were carefully chosen (Fig. 2 and Fig. 6S in Ref. [19]), however these data are not sufficient for performing an univocal indexing.

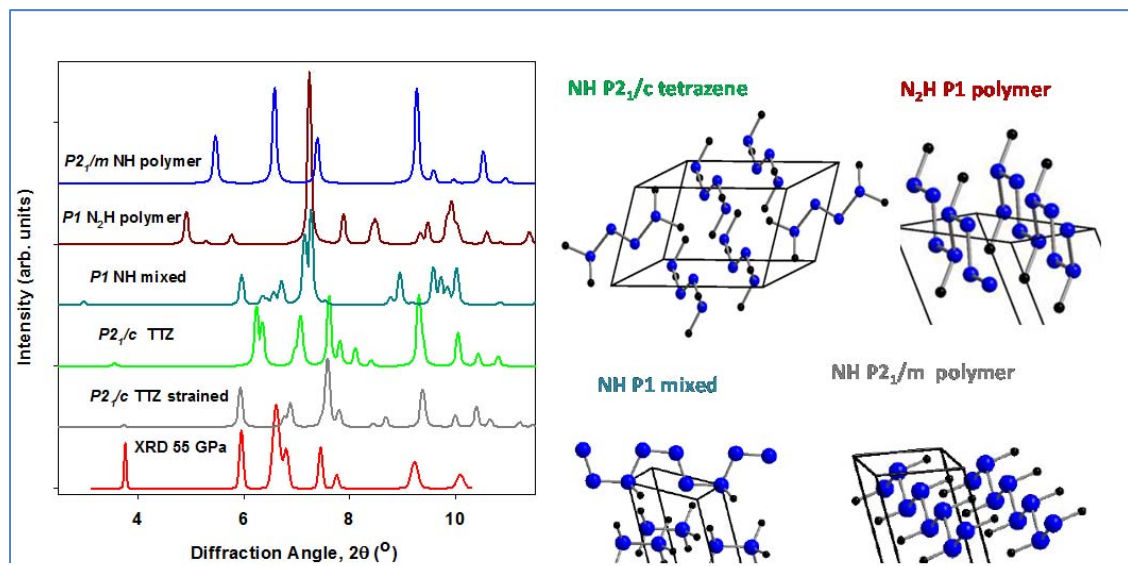

**Fig. 2 (color online). X-ray diffraction of a high-pressure phase at 55 GPa in comparison to the computed patterns of the theoretically predicted $N_xH$ structures at 50 GPa, this work and Ref. [11]. The x-ray wavelength was 0.2893 Å. The right panels show the projections of the structures illustrating the presence of molecules and/or indefinitely long N-N chains.**

Further insight into possible crystal structure can be given based on theoretical calculations with the predictive power [24, 25]. Polymeric structures in the N-H system become more stable at high pressures (*e.g.*, Ref. [10]). Search for the most stable structures with a variable N-H composition using evolutionary and random algorithms (USPEX [26] and AIRSS [25] codes) yielded a variety of different compositions and structures, the majority of which are quite complex and contain a large number of atoms. The detailed report of the study performed using USPEX will be published elsewhere (Qian *et al.*, unpublished).

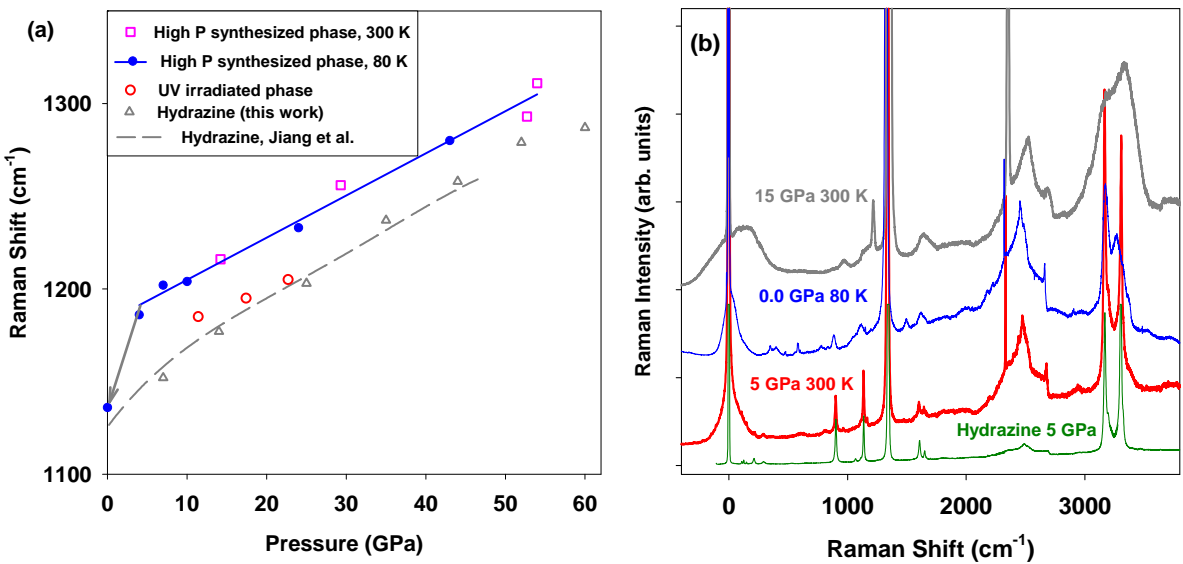

**Fig. 3 (color online). Metastability of the synthesized at high pressure polymeric material on the pressure decrease: (a) N-N stretching mode frequency; symbols: this work –different experiments; lines – guides to the eye; the arrow shows an abrupt change in frequency at unloading to 0 GPa at 80 K; the dashed line is the hydrazine data from Ref. [20]. (b) Raman spectra: the bottom trace corresponds to hydrazine spectra at 5 GPa.**

The detailed examination of the theoretically proposed $N_xH$ structures and their enthalpies show that above 40 GPa, in addition to ammonia $NH_3$, the compounds with compositions NH and $NH_2$ become thermodynamically stable (Fig. 4). Two $N_2H$ structures that are very close in enthalpy (*P1* and *P2₁/c*) consist of $N_4H_2$ buckled infinite chains (Fig. 2). There are two very close in enthalpy structures for NH compounds: (*P1* and *P2₁/c*) (Figs. 2, 4 and Fig. S7 in Ref. [19]). Below 55 GPa the stable one is *P2₁/c* , the structure (Fig. 2), which is similar to *trans*-tetrazene (TTZ) $H_2N$-N-N-$NH_2$ predicted in Ref. [11]. The structure of the second one (called mixed) is more complex: it consists of interchanging layers of $N_4H$ buckled infinite polymeric chains and $N_2H_5$ quasi-molecules (Fig. 2). The comparison of the experimental and theoretically predicted XRD shows a similarity for $P2_1/c$ tetrazene structure (Fig. 2), but much less resemblance to other structures predicted here and the $P2_1/m$ structure of Ref. [11]. Moreover, we find that a small strain of $P2_1/c$ tetrazene structure (within 3.5% in lattice parameters) would make the agreement better (Fig. 2).

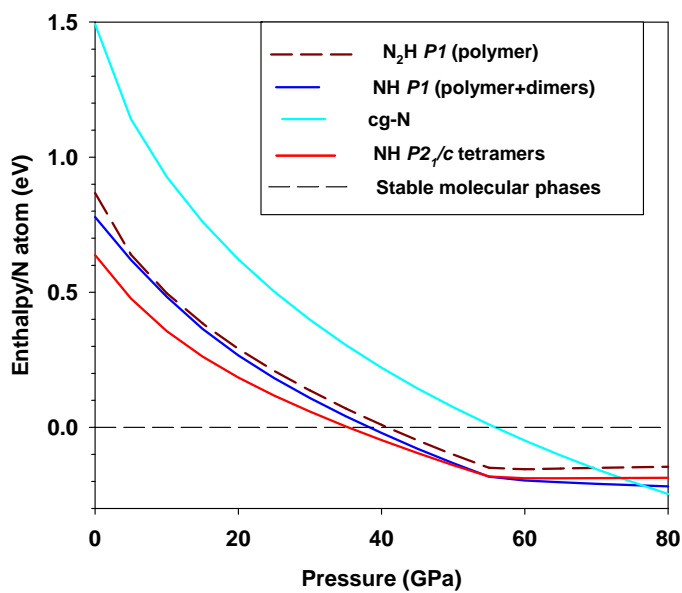

**Fig. 4 (color online). The enthalpies of the most stable hydronitrogens and cg-N determined theoretically as a function of pressure (relative to a mixture of molecular phases).**

A detailed comparison of the experimental and theoretical Raman and IR spectra (Fig. 1 and Fig. S8 in Ref. [19]) also suggest that $P2_1/c$ tetrazene is the compound synthesized at high pressures. Based on this comparison, we can safely rule out the $N_2H$ polymeric compound and NH mixed polymer-dimer structure. However, taken into account a poor agreement of the diffraction data, we cannot completely rule out structures which are more complex or even slightly different from the predicted. Moreover, the material may be a mixture of several phases including those which we have predicted here. This is further supported by the proximity of the enthalpies (Fig. 4). We can further speculate that the structural disagreement (Fig. 2) may be attributed to the presence of defects due to frustration of the N-N bond directions in polymeric chains, thus making various types of -N-N- oligomers to occur simultaneously. The presence of the $N_2$ inclusions or defects, which we have documented (Fig. 1), can also affect the stress and structural state of hydronitrogens.

The high-pressure phase has a large range of (meta)stability as evident from Raman, IR absorption, and X-ray diffraction data which we obtained on pressure release (Fig. 3). At room temperature the new phase is stable down to 15 GPa. At lower pressures, the material experiences decomposition, and the Raman and the IR spectra of the reaction products identify them as hydrazine (which we studied in a separate control high-pressure run, see also Ref. [20]) and molecular nitrogen, $N_2$. We have also performed a separate Raman experiment on unloading the high pressure phase at low temperatures (80 K), with the results showing its remarkable stability down to ambient pressure. The Raman bands (especially the N-H stretch modes) at low pressures and 80 K are substantially narrower than at high pressure/temperatures. Moreover, some bands develop a structure with several components, but, nevertheless, the major spectral components remain supporting the conclusion about the metastability of the high-pressure phase (Fig. 3), although small traces of molecular $N_2$, $H_2$, and hydrazine are also observed. The Raman band, which is close in frequency to $\nu_2$ $N_2$ vibron band as well a low-frequency lattice mode (Fig. 1c), rapidly drops in intensity at 0 GPa (80 K), indicating that these bands stem from interstitial and/or phase separated molecular $N_2$. The high-pressure phase rapidly decomposes (observed visually) forming again hydrazine (verified by Raman spectroscopy) on heating above 130 K.

It is remarkable that our experiment shows the transformation at pressures (47 GPa), which is only slightly higher than the predicted theoretically thermodynamic boundary (40 GPa, Fig. 4). The reverse transformation was found at about 15 GPa, suggesting that the experimental equilibrium transition pressure is near 31 GPa. Similarly, theory suggests that cgN is stable above 58 GPa (Fig. 4, see also Ref. [27]), which is lower than the experimental pressure (110 GPa) [2]. We speculate that $N_2$ and $H_2$ molecules, forming a low-pressure inclusion compound (this work and Ref. [16]), are kinetically prompt for dissociation and formation of backbone NH chains even at room temperature. Here we found that these molecules interact with each other in an unusual manner (see also Ref. [16]) revealed in Raman vibron spectra (Fig. 2S in Ref. [19]). Moreover, this chemical transformation can be initiated by optical photon irradiation, promoting the molecules to the excited states. The energy difference between the oligomeric and molecular states increases sharply below 40 GPa and it accounts approximately 60 % of that of cgN at ambient pressure (Fig. 4).

Our combined experimental and theoretical study of $N_xH$ system provides a first unambiguous evidence of novel mechano- and photochemistry which results in formation of polymeric/ oligomeric N-N backbone compounds that are recoverable to ambient pressure. First principles DFT (Fig. 4) and simplistic bond energy calculations (Table S1 in Ref. [19]) show that materials synthesized here possess an energy yield up to 61 % of that of cubic gauche nitrogen depending on the length of the -N-N- chains. This finding enables search for technologically relevant synthesis techniques of such materials which holds a promise to be the choice for fuel of the future.

We thank Michael Armstrong for valuable comments and X.-J. Chen and Z. Zhi for supporting this work in China. We acknowledge support from the Army Research Office, DARPA, NSF EAR and NSF EAR/IF, and EFREE, a BES-EFRC center at Carnegie. X-ray diffraction experiments were performed at GeoSoilEnviroCARS (Sector 13), Advanced Photon Source (APS), Argonne National Laboratory and Petra III, DESY, Hamburg, Germany. GeoSoilEnviroCARS is supported by the National Science Foundation - Earth Sciences (EAR-1128799) and Department of Energy - Geosciences (DE-FG02-94ER14466). Use of the Advanced Photon Source was supported by the U. S. Department of Energy, Office of Science, Office of Basic Energy Sciences, under Contract No. DE-AC02-06CH11357. PETRA III at DESY is a member of the Helmholtz Association (HGF). We thank the National Science Foundation (EAR-1114313, DMR-1231586), DARPA (Grants No. W31P4Q1210008 and No. W31P4Q1310005), the Government (No. 14.A12.31.0003) and the Ministry of Education and Science of Russian Federation (Project No. 8512) for financial support, and Foreign Talents Introduction and Academic Exchange Program (No. B08040). Calculations were performed on XSEDE facilities and on the cluster of the Center for Functional Nanomaterials, Brookhaven National Laboratory, which is supported by the DOE-BES under contract no. DE-AC02-98CH10086. The research leading to these results has received funding from the European Community's Seventh Framework Programme (FP7/2007-2013) under grant agreement n° 312284.

1. W. B. Hubbard, W. J. Nellis, A. C. Mitchell, N. C. Holmes, S. S. Limaye and P. C. McCandless, Science **253**, 648 (1991).
2. M. Eremets, I. Trojan, A. Gavriliuk and S. Medvedev, in *Static Compression of Energetic Materials*, ed. S. M. Peiris and G. J. Piermarini (Springer Berlin Heidelberg, 2008), pp. 75-97.
3. M. I. Eremets, A. G. Gavriliuk, I. A. Trojan, D. A. Dzivenko and R. Boehler, Nat. Mater. **3**, 558 (2004).
4. P. C. Samartzis and A. M. Wodtke, International Reviews in Physical Chemistry **25** (4), 527-552 (2006).
5. B. Hirshberg, R. B. Gerber and A. I. Krylov, Nat. Chem. **6**, 52 (2014).
6. K. O. Christe, W. W. Wilson, J. A. Sheehy and J. A. Boatz, Ang. Chem. Int. Ed. **38**, 2004 (1999).
7. Y.-C. Li, C. Qi, S.-H. Li, H.-J. Zhang, C.-H. Sun, Y.-Z. Yu and S.-P. Pang, J. Am. Chem. Soc. **132**, 12172 (2010).
8. T. M. Klapötke and D. G. Piercey, Inorg. Chem. **50**, 2732 (2011).
9. T. M. Klapötke, J. Stierstorfer and A. U. Wallek, Chem. of Mater. **20**, 4519 (2008).
10. V. V. Brazhkin and A. G. Lyapin, Nat. Mater. **3**, 497 (2004).
11. A. Hu and F. Zhang, J. Phys.: Condens. Matter **23**, 022203 (2011).
12. M. Ceppatelli, R. Bini and V. Schettino, Proc. Nat. Acad. Sci. U.S.A. **106**, 11454 (2009).
13. W. L. Mao, H.-k. Mao, Y. Meng, P. J. Eng, M. Y. Hu, P. Chow, Y. Q. Cai, J. Shu and R. J. Hemley, Science **314**, 636 (2006).
14. D. Chelazzi, M. Ceppatelli, M. Santoro, R. Bini and V. Schettino, Nat. Mater. **3**, 470 (2004).
15. J. A. Ciezak, T. A. Jenkins and R. J. Hemley, AIP Conf. Proc. **1195**, 1291 (2009).
16. M. Kim and C.-S. Yoo, J. Chem. Phys. **134**, 044519 (2011).
17. J. J. Gilman, Science **274**, 65 (1996).
18. C. A. M. Seidel and R. Kuhnemuth, Nat. Nano **9**, 164 (2014).
19. See Supplemental Material at http://link.aps.org/supplemental/10.1103/PhysRevLett. for additional details.

20. S. Jiang, X. Huang, D. Duan, S. Zheng, F. Li, X. Yang, Q. Zhou, B. Liu and T. Cui, J. Phys. Chem. C **118**, 3236 (2014).
21. T. Shimanouchi, Pure Appl. Chem. **36**, 93 (1973).
22. V. Schettino and R. Bini, Phys. Chem. Chem. Phys. **5**, 1951 (2003).
23. E. Gregoryanz, A. F. Goncharov, C. Sanloup, M. Somayazulu, H.-k. Mao and R. J. Hemley, J. Chem. Phys. **126**, 184505 (2007).
24. A. R. Oganov and C. W. Glass, J. Chem. Phys. **124** (24), 244704 (2006).
25. C. J. Pickard and R. J. Needs, J. of Phys.: Cond. Mat. **23** (5), 053201 (2011).
26. A. O. Lyakhov, A. R. Oganov, H. T. Stokes and Q. Zhu, Computer Physics Communications **184**, 1172 (2013).
27. C. J. Pickard and R. J. Needs, Phys. Rev. Lett. **102**, 125702 (2009).
28. A. F. Goncharov, P. Beck, V. V. Struzhkin, B. D. Haugen and S. D. Jacobsen, PEPI **174**, 24 (2009).
29. V. B. Prakapenka, A. Kubo, A. Kuznetsov, A. Laskin, O. Shkurikhin, P. Dera, M. L. Rivers and S. R. Sutton, High Press. Res. **28**, 225 (2008).
30. A. R. Oganov, Y. M. Ma, A. O. Lyakhov, M. Valle and C. Gatti, Rev. Mineral. Geochem. **71**, 271 (2010).
31. J. P. Perdew, K. Burke and M. Ernzerhof, Phys. Rev. Lett. **77**, 3865 (1996).
32. G. Kresse and J. Furthmüller, Comp. Mater. Sci. **6**, 15 (1996).
33. S. J. Clark, M. D. Segall, C. J. Pickard, P. J. Hasnip, M. J. Probert, K. Refson and M. C. Payne, Zeit. Kristallogr. **220**, 567 (2005).

# Backbone N$_x$H Compounds at High Pressures

Alexander F. Goncharov, Nicholas Holtgrewe, Guangrui Qian, Chaohao Hu, Artem R. Oganov, M. Somayazulu, E. Stavrou, Chris J. Pickard, Adam Berlie, Fei Yen, M. Mahmood, S. S. Lobanov, Z. Konôpková, V. B. Prakapenka

**Materials and Methods:**

Molecular nitrogen and hydrogen were mixed in a high-pressure cylinder at 5-10 MPa. The sample composition was estimated based on the initial partial pressures. Based on this estimation, our samples of N$_x$H had composition x in the range of 1.5 to 0.67. Well homogenized over several days molecular mixtures were loaded in a symmetric diamond anvil cell at 200 MPa at room temperature. Then the pressure was increased slowly and the samples were probed by *in situ* micro Raman and optical spectroscopy (including infrared) and synchrotron x-ray diffraction (see below). The reaction products were unloaded at 297 and 80 K monitoring the sample state with the same techniques.

Raman studies were performed using 488 and 532 nm lines of a solid-state laser. The laser probing spot dimension was 4 μm. Raman spectra were analyzed with a spectral resolution of 4 cm$^{-1}$ using a single-stage grating spectrograph equipped with a CCD array detector. Optical absorption spectra in the visible and near IR spectral ranges were measured using an all-mirror custom microscope system coupled to a grating spectrometer equipped with a CCD detector [28]. X-ray diffraction measurements were performed at the undulator XRD beamline at GeoSoilEnviroCARS, APS, Chicago [[29]] and Extreme Conditions Beamline P02.2 at DESY (Germany). The X-ray probing beam size was 2-5 μm.

UV and visible multi-photon absorption was performed with ultra-short pulses (370, 580, & 720 nm) provided by an optical parametric amplifier (Coherent OPerA) with input pulses generated from a femtosecond oscillator combined with chirped pulse amplification (Coherent Mantis and Legend Elite). The radiation was focused in a spot of approximately 10 μm in diameter with an average power of 200μW and the sample was irradiated for a few hours prior to analysis.

Predictions of stable phases were done using the USPEX code in the variable-composition mode [30]. The first generation of structures was produced randomly and the succeeding generations were obtained by applying heredity, atom transmutation, and lattice mutation operations, with probabilities of 60%, 10% and 30%, respectively. 80% non-identical structures of each generation with the lowest enthalpies were used to produce the next generation. All structures were relaxed using density functional theory (DFT) calculations within the Perdew-Burke-Ernzerhof (PBE) [31], as implemented in the VASP code [32]. Further Ab Initio Random Structures Searches (AIRSS) [25] were performed using the CASTEP code [33] with very similar results, and producing the structure which shows good agreement with the experimental diffraction data.

**Supplemental Figures**

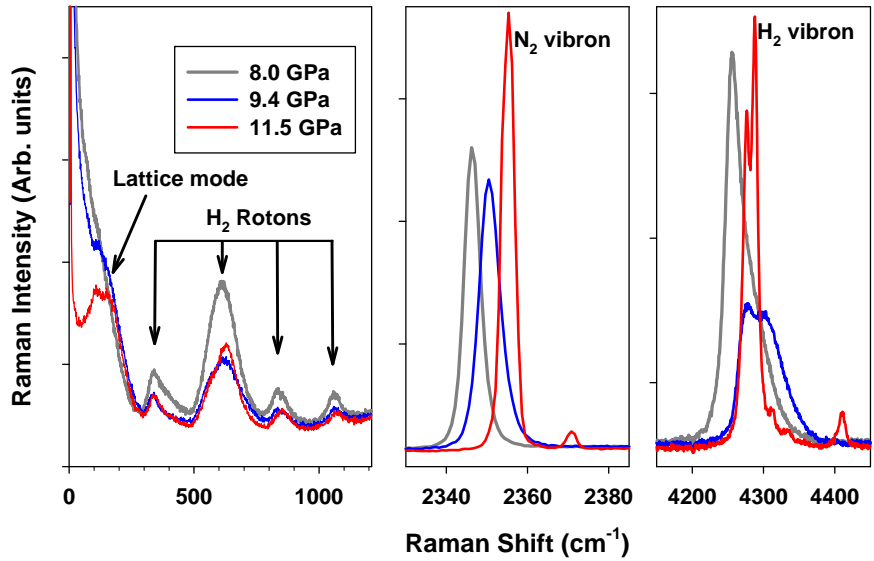

**Fig. S1**. Raman spectra of the $N_2$-$H_2$ mixture through solidification. In a crystalline phase (at 11.5 GPa) clear overdamped lattice modes are seen near 150 cm$^{-1}$, there are minor changes on the $H_2$ rotational modes, and the vibron modes split and develop sidebands.

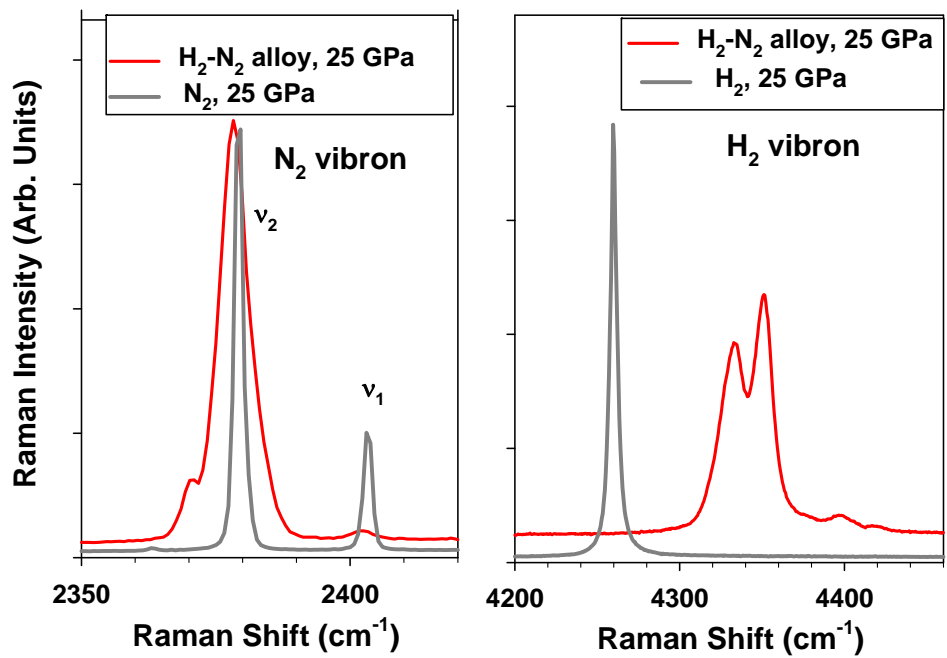

**Fig. S2**. Raman vibron spectra of the $N_2$-$H_2$ mixture in comparison to those of pure $N_2$ and $H_2$ at 25 GPa. The Raman spectra of the $N_2$-$H_2$ compound show more complex vibron structure and broader peaks.

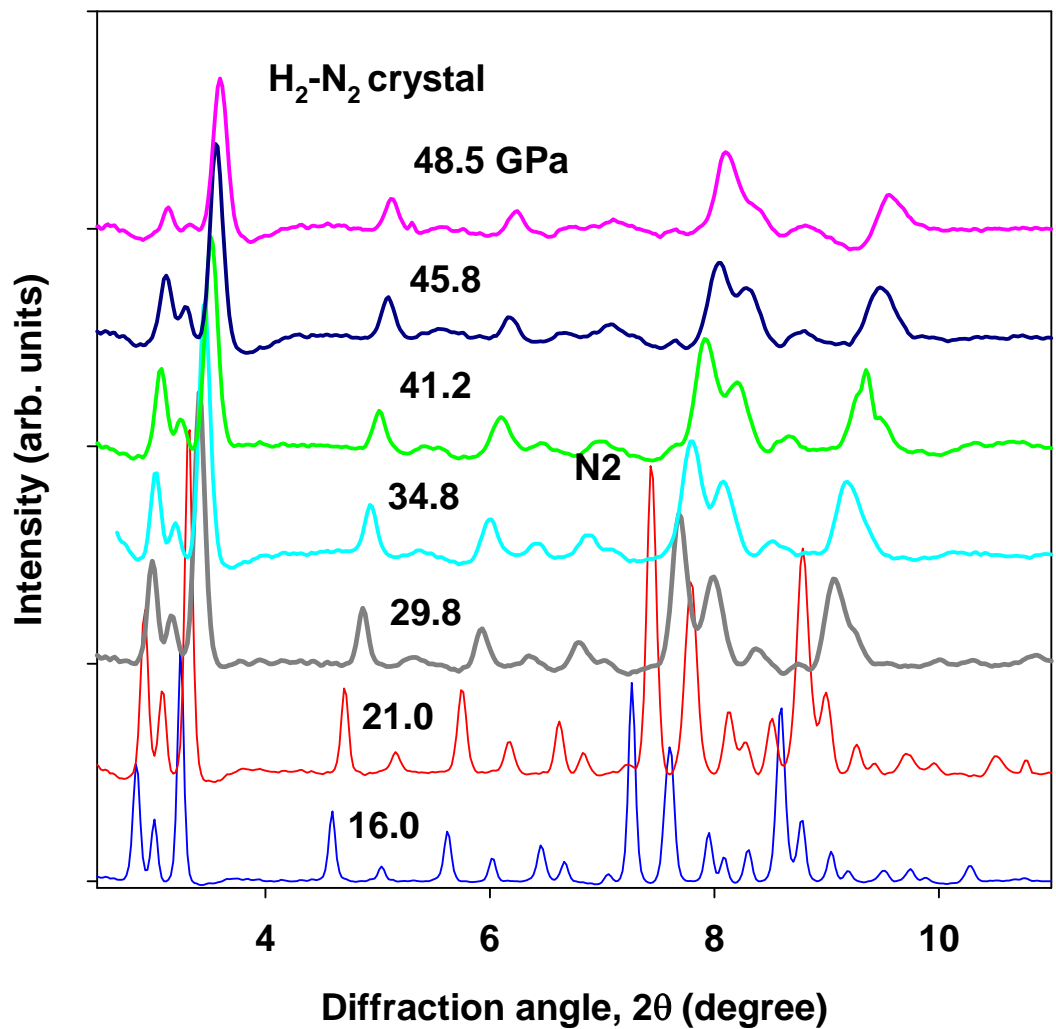

**Fig. S3**. X-ray diffraction patterns of an $H_2$-$N_2$ van-der-Waals crystal as a function of pressure. The x-ray wavelength was 0.3344 Å.

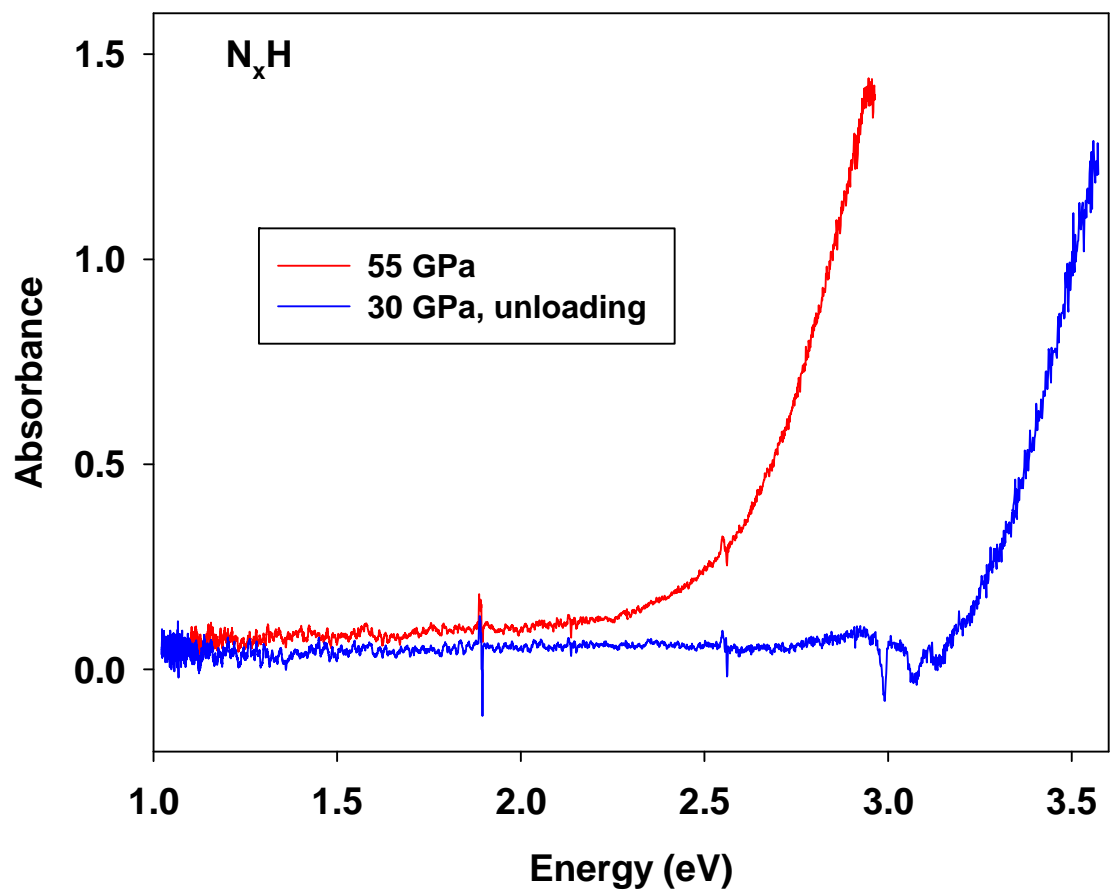

**Fig. S4**. Optical absorption spectra after the transformation to a high-pressure oligomeric phase at 55 GPa and at 30 GPa on the pressure decrease.

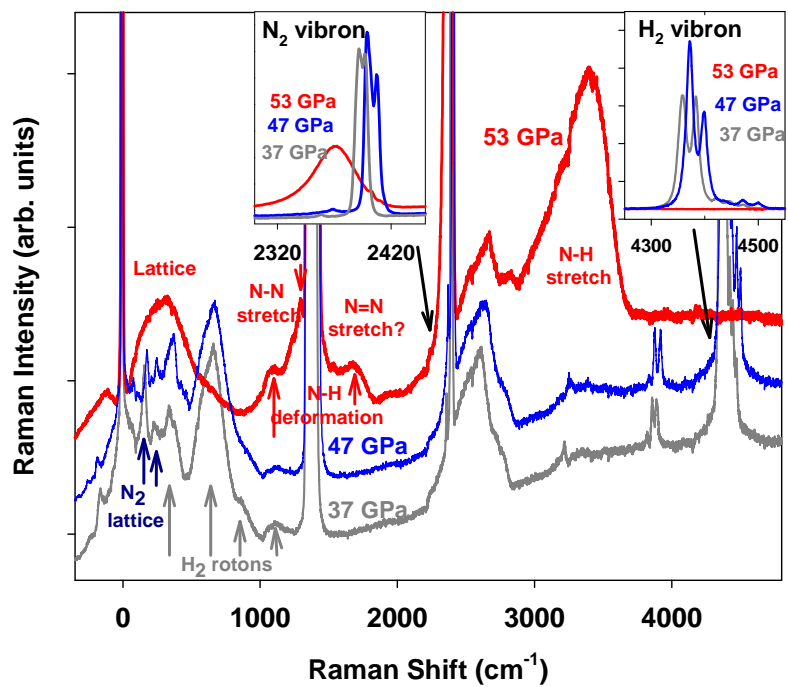

**Fig. S5**. Raman spectra at increasing pressures and with time evolution. The spectrum at 53 GPa has been measured 4 days after that measured at 47 GPa; the pressure was gradually increased during this time by small steps. The insets show details of the $N_2$ and $H_2$ vibron spectra behavior.

**(a)**

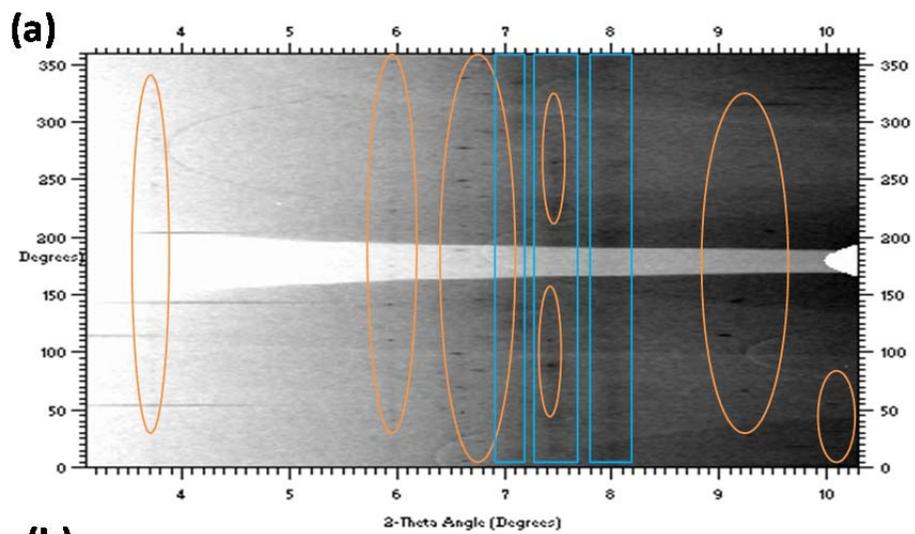

**(b)**

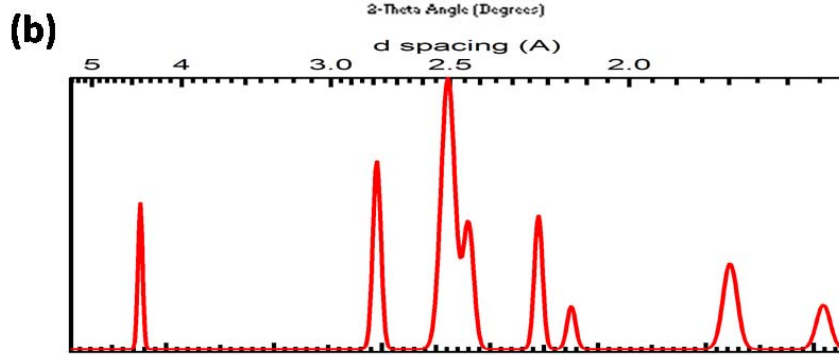

**(c)**

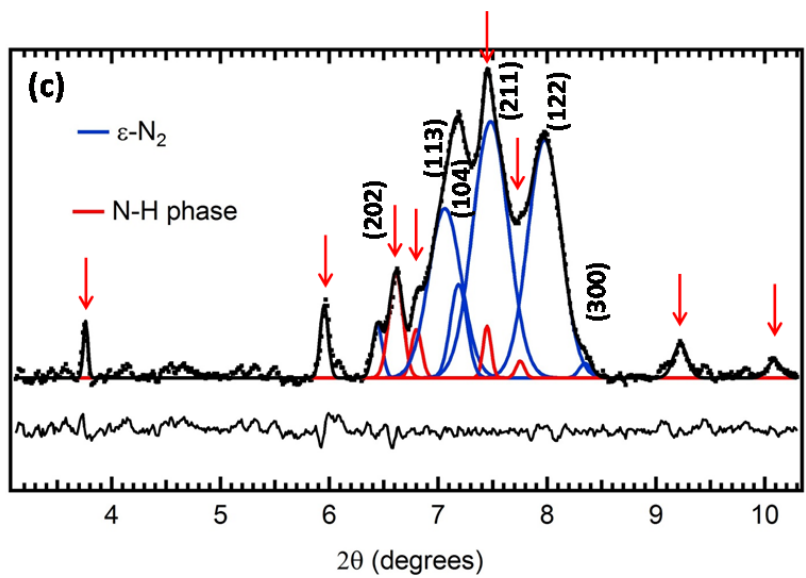

**Fig. S6**. X-ray diffraction pattern at 55 GPa after the transformation to the oligomeric phase. (a) 2D x-ray diffraction (cake), the ovals and rectangulars show the areas with diffraction of the new phase and molecular nitrogen, respectively; (b) extracted x-ray diffraction of a synthesized phase; (c) the deconvolution of peaks of molecular nitrogen (the main peaks are indexed) and oligomeric phase (marked by red arrows) in the integrated pattern.

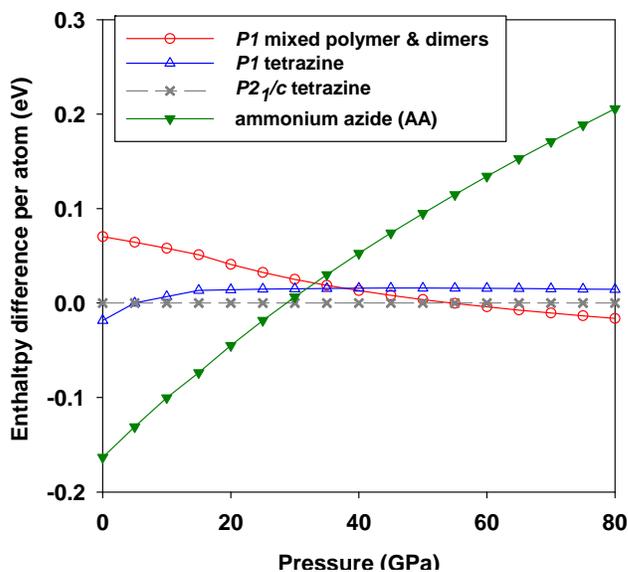

**Fig. S7**. The enthalpies of the most stable theoretically proposed in this work structures of hydronitrogens. The enthalpy of ammonium azide is shown for comparison. The transition to a tetrazine structure occurs near 27 GPa (cf. 36 GPa in Ref. [11]).

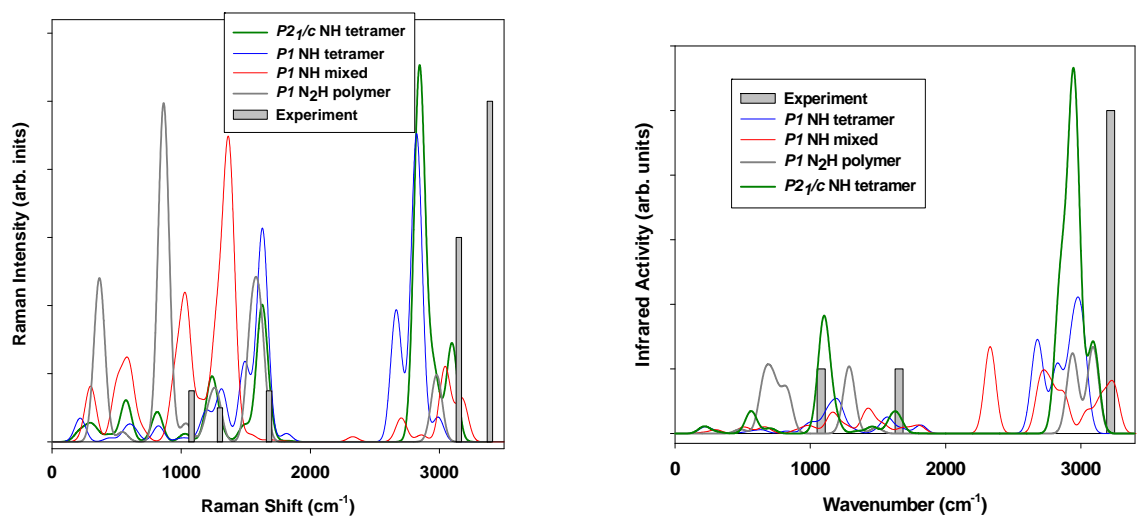

**Fig. S8**. Theoretically computed at 50 GPa Raman and IR spectra of the most stable hydronitrogens in comparison with the experimental data (at 55 GPa) for the compound synthesized at 47 GPa.

Table S1. Bond energies (kJ/mole) referred to N-N and H-H bond energies in diatomics.

| Molecule | Stucture | Reaction[1] | $\Delta H_{reaction}$ (kJ/mol)[2] |
|---|---|---|---|
| N-H Polymer[3] | 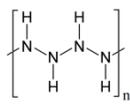 | $(NH)_4 \rightarrow 2N_2 + 2H_2$ | −134 |
| Tetrazene | 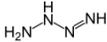 | $(NH)_4 \rightarrow 2N_2 + 2H_2$ | −113 |
| Ammonium Azide | 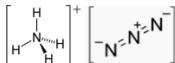 | $(NH)_4 \rightarrow 2N_2 + 2H_2$ | −92 |
| Hydrazine | 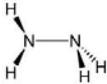 | $2N_2H_4 \rightarrow 2N_2 + 4H_2$ | −47.5 |
| Ammonia | 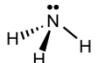 | $4NH_3 \rightarrow 2N_2 + 6H_2$ | +39 |
| Cubic Gauche Nitrogen | 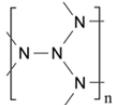 | $N_4 \rightarrow 2N_2$ | −220.5 |

[1]All reactions normalized to 1 Nitrogen atom.
[2]Calculations based on gas phase bond enthalpies only, enthalpy of sublimation (or vaporization) not taken into account.
[3]The synthesized material in this work is either a tetrazene or a mixed polymer - dimer.